\documentstyle[a4,12pt]{article}
\begin{document}

\begin{flushright}
{\bf Preprint LMU-98-10} \\
{September 1998}
\end{flushright}

\vspace{0.2cm}

\begin{center}
{\large\bf Measuring $CP$ Violation and Testing Factorization \\
in $B_d\rightarrow D^{*\pm}D^{\mp}$ and $B_s
\rightarrow D^{*\pm}_sD^{\mp}_s$ Decays}
\end{center}

\vspace{1.2cm}

\begin{center}
{\bf Zhi-zhong Xing}\footnote{
E-mail: Xing$@$hep.physik.uni-muenchen.de }\\
{\sl Sektion Physik, Universit${\sl\ddot a}$t M${\sl\ddot
u}$nchen, Theresienstrasse 37A, 80333 M${\sl\ddot u}$nchen, Germany}
\end{center}

\vspace{2cm}

\begin{abstract}
We show that the $CP$-violating
phases $\beta$ and $\beta'$
can respectively be determined from the
time-dependent measurements of 
$B_d \rightarrow D^{*\pm}D^{\mp}$ and $B_s \rightarrow
D^{*\pm}_sD^{\mp}_s$ decays, whose final states are
non-$CP$ eigenstates. 
The penguin contributions to the mixing-induced
interference quantities are expected to be about
$4\%$ or smaller. 
We also point out two observables of $O(1)$,
which are pure functions of
decay constants and form-factors, for
a clean test of the factorization hypothesis in
neutral-$B$ decays into two heavy charmed mesons.
\end{abstract}

\newpage
\section{Introduction}

In the first-round experiments of $B$-meson factories, the 
measurement of $B_d \rightarrow D^+D^-$, $D^{*+}D^-$, 
$D^+D^{*-}$ and $D^{*+}D^{*-}$ decay modes will be available.
These channels are expected to 
have fairly large branching ratios
(of order $10^{-4}$ to $10^{-3}$) and involve the $CP$-violating phase
\begin{equation}
\beta \; \equiv \; \arg \left (- \frac{V^*_{tb}V_{td}}
{V^*_{cb}V_{cd}} \right ) \; ,
\end{equation}
where $V_{ij}$ (for $i=u,c,t$ and $j=d,s,b$) are elements of
the Cabibbo-Kobayashi-Maskawa (CKM) matrix. A determination of
$\beta$ from $CP$ asymmetries of $B_d \rightarrow
D^{(*)+}D^{(*)-}$ will be useful, not only to cross-check the
extraction of $\beta$ from $B_d \rightarrow J/\psi K_S$, but
also to shed some light on the relevant penguin effect and final-state
interactions \cite{Xing97}. 
In addition, it should be important to test whether the
factorization approximation, which works well
for a variety of $B$ transitions into one light and one heavy
mesons such as $D\pi$ and $D\rho$ \cite{Stech},
remains valid for the $B$ decays into two heavy charmed mesons.

In the $B^0_s$-$\bar{B}^0_s$ system, the similar decay channels
are $B_s \rightarrow D^+_sD^-_s$, $D^{*+}_sD^-_s$, $D^+_sD^{*-}_s$
and $D^{*+}_sD^{*-}_s$. The branching ratios of these
transitions are expected to be of order $10^{-2}$, 
therefore they are rapidly accessible at a high-energy
$B$ factory. The $CP$-violating phase
\begin{equation}
\beta' \; \equiv \; \arg \left (-\frac{V^*_{tb}V_{ts}}
{V^*_{cb}V_{cs}} \right ) \; , 
\end{equation}
which is negligibly small in the standard model
\footnote{Note that this result is independent of any
specific parametrization of the CKM matrix. Taking
into account 
$V^*_{ub}V_{us} + V^*_{cb}V_{cs} + V^*_{tb}V_{ts} =0$
and $|V_{cb}V_{cs}| \approx |V_{tb}V_{ts}| \gg
|V_{ub}V_{us}|$ \cite{PDG98}, one arrives straightforwardly at 
$|\beta'| \leq \arctan | (V_{ub}V_{us})/
(V_{cb}V_{cs}) | \sim 1^{\circ}$.},
will be determinable from $CP$ asymmetries of
$B_s \rightarrow D^{(*)+}_sD^{(*)-}_s$. The measured
value of $|\beta'|$, if remarkably larger than $1^{\circ}$,
would clearly imply the existence
of new physics in $B^0_s$-$\bar{B}^0_s$ mixing. Whether the 
factorization approximation works well for such decay modes
is also an open question.

As neutral-$B$ decays into $CP$ eigenstates 
($B_d\rightarrow D^+D^-$, etc.) have been studied
in some detail \cite{Xing97,Bigi}, 
the present note concentrates only on $B_d \rightarrow 
D^{*\pm}D^{\mp}$ and $B_s \rightarrow D^{*\pm}_sD^{\mp}_s$
decays, whose final states are non-$CP$ eigenstates.
These decay modes, 
in comparison with $B_d \rightarrow D^{\pm}\pi^{\mp}$ 
\cite{Xing95},
have an obvious advantage for the study of 
possible $CP$ violation: the amplitudes of
$B^0$ and $\bar{B}^0$ into a common $DD^*$ state
are comparable in magnitude and therefore large interference
between them becomes posssible.
We show that
the weak phases $\beta$ and $\beta'$ 
can respectively be extracted 
from the time-dependent measurements of $B_d\rightarrow
D^{*\pm}D^{\mp}$ and $B_s\rightarrow D^{*\pm}_sD^{\mp}_s$
decays.
A large $CP$ asymmetry is 
expected to appear in the $B_d$ channels within the standard
model. 
The penguin contributions to the mixing-induced interference
quantities in these decay modes are estimated to be about $4\%$
or smaller.
Thus it is a good approximation to neglect the
relevant penguin effects. 
We also show that the time-integrated measurements
allow us to  determine two parameters of $O(1)$, which are pure
functions of decay constants and form-factors in the
factorization approximation:
\begin{eqnarray}
\zeta_d & = & \frac{f_D}{f_{D^*}}
\cdot \frac{A^{B_dD^*}_0(m^2_D)}{F^{B_dD}_1 (m^2_{D^*})} \; , \nonumber \\
\zeta_s & = & \frac{f_{D_s}}{f_{D^*_s}}
\cdot \frac{A^{B_sD^*_s}_0(m^2_{D_s})}{F^{B_sD_s}_1 (m^2_{D^*_s})} \; .
\end{eqnarray}
Comparing the experimental and theoretical results of
$\zeta_d$ or $\zeta_s$
will provide a clean test of the factorization
hypothesis for neutral-$B$ decays into two heavy charmed mesons.
Finally we give some brief comments on possible effects of new
physics on $B^0$-$\bar{B}^0$ mixing and $CP$ asymmetries.

\section{$\beta$, $\beta'$ and $CP$ asymmetries}

The transitions $B^0_d \rightarrow D^{*\pm}D^{\mp}$ and
$B^0_s\rightarrow D^{*\pm}_s D^{\mp}_s$ can occur through
both tree-level and loop-induced (penguin) quark diagrams,
The penguin contribution to the
overall amplitude of each decay mode
is tentatively neglected, as its
magnitude is small enough in comparison with that of the
tree-level amplitude (see the next section for a detailed
discussion about the penguin effect). With the help of
this good approximation, one may define two
interference quantities between decay amplitudes and 
$B^0_d$-$\bar{B}^0_d$ mixing:
\begin{eqnarray}
\lambda_{D^{*+}D^-} & \equiv & \frac{q^{~}_d}{p^{~}_d} 
\cdot \frac{A(\bar{B}^0_d\rightarrow D^{*+}D^-)}
{A(B^0_d\rightarrow D^{*+}D^-)} \; =\;
\frac{V^*_{tb}V_{td}}{V_{tb}V^*_{td}}\cdot
\frac{V_{cb}V_{cd}^*}{V^*_{cb}V_{cd}} 
~ \zeta_d ~ e^{{\rm i}\delta_d} \; =\;
\zeta_d ~ e^{{\rm i} (\delta_d + 2\beta)} \; , \nonumber \\
\bar{\lambda}_{D^{*-}D^+} & \equiv & \frac{p^{~}_d}{q^{~}_d} 
\cdot \frac{A(B^0_d\rightarrow D^{*-}D^+)}
{A(\bar{B}^0_d\rightarrow D^{*-}D^+)} \; =\;
\frac{V_{tb}V^*_{td}}{V^*_{tb}V_{td}}\cdot
\frac{V^*_{cb}V_{cd}}{V_{cb}V^*_{cd}} 
~ \zeta_d ~ e^{{\rm i}\delta_d} \; =\;
\zeta_d ~ e^{{\rm i} (\delta_d - 2\beta)} \; ,
\end{eqnarray}
where $q^{~}_d/p^{~}_d = (V^*_{tb}V_{td})/(V_{tb}V^*_{td})$
describes the $B^0_d$-$\bar{B}^0_d$ mixing
phase in the box-diagram approximation, $\zeta_d$ and
$\delta_d$ denote
the ratio of the real hadronic matrix elements and
the strong phase difference between
$\bar{B}^0_d\rightarrow D^{*+}D^-$ and $B^0_d\rightarrow
D^{*+}D^-$. 
For the similar transitions $B^0_s$ vs $\bar{B}^0_s \rightarrow
D^{*+}_sD^-_s$ and $D^{*-}_sD^+_s$, we have
\begin{eqnarray}
\lambda_{D^{*+}_sD^-_s} & \equiv & \frac{q^{~}_s}{p^{~}_s} 
\cdot \frac{A(\bar{B}^0_s\rightarrow D^{*+}_sD^-_s)}
{A(B^0_s\rightarrow D^{*+}_sD^-_s)} \; =\;
\frac{V^*_{tb}V_{ts}}{V_{tb}V^*_{ts}}\cdot
\frac{V_{cb}V_{cs}^*}{V^*_{cb}V_{cs}} 
~ \zeta_s ~ e^{{\rm i}\delta_s} \; =\;
\zeta_s ~ e^{{\rm i} (\delta_s + 2\beta')} \; , \nonumber \\
\bar{\lambda}_{D^{*-}_sD^+_s} & \equiv & \frac{p^{~}_s}{q^{~}_s} 
\cdot \frac{A(B^0_s\rightarrow D^{*-}_sD^+_s)}
{A(\bar{B}^0_s\rightarrow D^{*-}_sD^+_s)} \; =\;
\frac{V_{tb}V^*_{ts}}{V^*_{tb}V_{ts}}\cdot
\frac{V^*_{cb}V_{cs}}{V_{cb}V^*_{cs}} 
~ \zeta_s ~ e^{{\rm i}\delta_s} \; =\;
\zeta_s ~ e^{{\rm i} (\delta_s - 2\beta')} \; ,
\end{eqnarray}
in which the notations are self-explanatory.
These quantities consist of both weak and strong phases
and will affect the decay rates in a significant way.
Their imaginary parts are of particular interest for
the study of $CP$ violation:
\begin{eqnarray}
{\rm Im} \lambda_{D^{*+}D^-} & = & \zeta_d ~
\sin (\delta_d + 2\beta) \; , \nonumber \\
{\rm Im} \lambda_{D^{*-}D^+} & = & \zeta_d ~
\sin (\delta_d - 2\beta) \; , \nonumber \\
{\rm Im} \lambda_{D^{*+}_sD^-_s} & = & \zeta_s ~
\sin (\delta_s + 2\beta') \; , \nonumber \\
{\rm Im} \lambda_{D^{*-}_sD^+_s} & = & \zeta_s ~
\sin (\delta_s - 2\beta') \; .
\end{eqnarray}
It should be noted, however, that 
${\rm Im}\lambda_f$ and ${\rm Im} \bar{\lambda}_{\bar{f}}$ (for
$f = D^{*+}D^-$ or $D^{*+}_s D^-_s$)
themselves are not $CP$-violating observables! Only
their difference
${\rm Im} (\lambda_f - \bar{\lambda}_{\bar{f}})$, which will
vanish for $\beta =\beta' =0$ or $\pi$, 
measures the $CP$ asymmetry.

The generic formulas for decay rates of neutral-$B$ into non-$CP$
eigenstates have been given in the literature (see, e.g.,
Refs. \cite{Bigi,Xing94}). For
the decay modes under consideration, their time-dependent
rates read as
\begin{eqnarray}
{\cal R} \left [ \stackrel{\langle - \rangle}{B^0_d}(t)
\rightarrow D^{*+}D^- \right ] 
& \propto & e^{-\Gamma_d t}
\left [\frac{1 +\zeta^2_d}{2} ~ \stackrel{\langle - \rangle}
{+} ~ \frac{1 -\zeta^2_d}{2} ~ \cos (x_d \Gamma_d t) 
\right . \nonumber \\
&  & \left . ~~~~~~ \stackrel{\langle + \rangle}{-} ~ 
\zeta_d ~ \sin (\delta_d + 2\beta) \sin (x_d \Gamma_d t)
\right ] \; , \nonumber \\
{\cal R} \left [ \stackrel{\langle - \rangle}{B^0_d}(t)
\rightarrow D^{*-}D^+ \right ] 
& \propto & e^{-\Gamma_d t}
\left [ \frac{1 +\zeta^2_d}{2} ~ \stackrel{\langle + \rangle}
{-} ~ \frac{1 -\zeta^2_d}{2} ~ \cos (x_d \Gamma_d t) 
\right . \nonumber \\
&  & \left . ~~~~~~ \stackrel{\langle - \rangle}{+} ~ 
\zeta_d ~ \sin (\delta_d - 2\beta) \sin (x_d \Gamma_d t)
\right ] \; ;
\end{eqnarray}
and
\footnote{Here we neglect the tiny effect
from the decay width difference between
two $B_s$ mass eigenstates. The latest theoretical calculation
\cite{Beneke}
predicts $\Delta \Gamma_s /\Gamma_s = 0.054^{+0.016}_{-0.032}$, 
which is negligibly small for our purpose.}
\begin{eqnarray}
{\cal R} \left [ \stackrel{\langle - \rangle}{B^0_s}(t)
\rightarrow D^{*+}_sD^-_s \right ] 
& \propto & e^{-\Gamma_s t}
\left [ \frac{1 +\zeta^2_s}{2} ~ \stackrel{\langle - \rangle}
{+} ~ \frac{1 -\zeta^2_s}{2} ~ \cos (x_s \Gamma_s t) 
\right . \nonumber \\
&  & \left . ~~~~~~ \stackrel{\langle + \rangle}{-} ~ 
\zeta_s ~ \sin (\delta_s + 2\beta') \sin (x_s \Gamma_s t)
\right ] \; , \nonumber \\
{\cal R} \left [ \stackrel{\langle - \rangle}{B^0_s}(t)
\rightarrow D^{*-}_sD^+_s \right ] 
& \propto & e^{-\Gamma_s t}
\left [ \frac{1 +\zeta^2_s}{2} ~ \stackrel{\langle + \rangle}
{-} ~ \frac{1 -\zeta^2_s}{2} ~ \cos (x_s \Gamma_s t) 
\right . \nonumber \\
&  & \left . ~~~~~~ \stackrel{\langle - \rangle}{+} ~ 
\zeta_d ~ \sin (\delta_s - 2\beta') \sin (x_s \Gamma_s t)
\right ] \; ,
\end{eqnarray}
where $x_d = 0.723 \pm 0.032$ and $x_s > 14.0$ (at the
$95\%$ confidence level) are $B^0_d$-$\bar{B}^0_d$
and $B^0_s$-$\bar{B}^0_s$ mixing parameters
\cite{PDG98}, and 
$\Gamma_d$ and $\Gamma_s$ denote decay widths. 
Note that the four decay rates in Eq. (7) or Eq. (8)
share a common factor, since the penguin effects in
these decays have been neglected.

The time-dependent measurements of $B_d\rightarrow 
D^{*\pm}D^{\mp}$ and $B_s \rightarrow D^{*\pm}_sD^{\mp}_s$,
as shown above, allow us to extract the weak phases $\beta$
and $\beta'$ up to a four-fold ambiguity:
\begin{eqnarray}
\sin^2 (2\beta) & = & \frac{1}{2} \left [(1 -S_+S_-)
~ \pm ~ \sqrt{(1-S^2_+) (1-S^2_-)} \right ] \; , 
\nonumber \\
\sin^2 (2\beta') & = & \frac{1}{2} \left [(1-S'_+S'_-)
~ \pm ~ \sqrt{(1-S^{'2}_+) (1-S^{'2}_-)} \right ] \; ,
\end{eqnarray}
where $S_{\pm} \equiv \sin (\delta_d \pm 2\beta)$
and $S'_{\pm} \equiv \sin (\delta_s \pm 2\beta')$.
Indeed only a two-fold ambiguity in $\sin(2\beta)$
and $\sin(2\beta')$ remains, because
$\sin (2\beta) >0$ and $\sin (2\beta') \sim 0$
are expected to be true within the standard model
\cite{PDG98}.
The strong phases $\delta_d$ and $\delta_s$ can 
also be determined from Eqs. (7) and (8) up to
a four-fold ambiguity:
\begin{eqnarray}
\sin^2 \delta_d & = & \frac{1}{2} \left [(1 +S_+S_-)
~ \pm ~ \sqrt{(1-S^2_+) (1-S^2_-)} \right ] \; , 
\nonumber \\
\sin^2 \delta_s & = & \frac{1}{2} \left [(1 +S'_+S'_-)
~ \pm ~ \sqrt{(1-S^{'2}_+) (1-S^{'2}_-)} \right ] \; .
\end{eqnarray}
If the final-state interactions were insignificant 
in these decay modes, 
$\delta_d$ and $\delta_s$ might not deviate too much from zero.
In this case, $S_+ \approx - S_-$ and $S'_+ \approx -S'_-$ 
would be a good approximation. 

The measurement of $S_{\pm}$ can be carried out at both
asymmetric $e^+e^-$ $B$-meson factories and high-luminosity
hadron machines. In contrast, a measurement of $S'_{\pm}$ 
is only available at hadronic $B$ factories. With all such
time-dependent measurements, the ratios of hadronic
matix elements $\zeta_d$
and $\zeta_s$ are determinable. Note that these two real
parameters can acturally be determined from the time-integrated
measurments at a symmetric $e^+e^-$ collider or at hadron
facilities of $B$ physics. For example, one may integrate
the decay rates in Eqs. (7) and (8) over $t \in [0, \infty )$
and obtain ratios of two decay rates as follows:
\begin{eqnarray}
\frac{{\cal R}(\bar{B}^0_{d,\rm phys} \rightarrow
D^{*+}D^-)}{{\cal R}(B^0_{d,\rm phys} \rightarrow
D^{*+}D^-)} & = &
\frac{\left (2+x^2_d \right ) \zeta^2_d ~ + ~ x^2_d}
{x^2_d \zeta^2_d ~ + ~ \left (2 + x^2_d \right )} \; \; ,
\nonumber \\
\frac{{\cal R}(\bar{B}^0_{s,\rm phys} \rightarrow
D^{*+}_sD^-_s)}{{\cal R}(B^0_{s,\rm phys} \rightarrow
D^{*+}_sD^-_s)} & = &
\frac{\left (2+x^2_s \right ) \zeta^2_s ~ + ~ x^2_s}
{x^2_s \zeta^2_s ~ + ~ \left (2 + x^2_s \right )} \; \; ,
\end{eqnarray}
where the subscript ``phys'' implies that the mixing effect
has been taken into account for $B^0_q$ and $\bar{B}^0_q$
mesons ($q=d$ or $s$).
Obviously the first relation of Eq. (11)
allows the determination of
$\zeta_d$. Once $x_s$ is known, $\zeta_s$ can be
extracted from the second relation of Eq. (11). If a
reliable calculation of $\zeta_s$ were possible, then
the magnitude of $x_s$ would reversely be obtained.

For the $B_d$ and $B_s$ decay modes discussed above, 
their time-dependent $CP$ asymmetries can be defined
as 
\begin{eqnarray}
{\cal A}_d(t) & \equiv & \frac{{\cal R}[B^0_d(t)
\rightarrow D^{*+}D^-] ~ - ~ {\cal R}[\bar{B}^0_d
(t)\rightarrow D^{*-}D^+]}
{{\cal R}[B^0_d(t)
\rightarrow D^{*+}D^-] ~ + ~ {\cal R}[\bar{B}^0_d
(t)\rightarrow D^{*-}D^+]} \; \; ,
\nonumber \\
{\cal A}_s(t) & \equiv & \frac{{\cal R}[B^0_s(t)
\rightarrow D^{*+}_sD^-_s] ~ - ~ {\cal R}[\bar{B}^0_s
(t)\rightarrow D^{*-}_sD^+_s]}
{{\cal R}[B^0_s(t)
\rightarrow D^{*+}_sD^-_s] ~ + ~ {\cal R}[\bar{B}^0_s
(t)\rightarrow D^{*-}_sD^+_s]} \; \; .
\end{eqnarray}
Taking Eqs. (7) and (8) into account, we explicitly obtain
\begin{equation}
{\cal A}_d(t) \; =\; \frac{-2\zeta_d \sin (2\beta)  
\cos\delta_d \sin (x_d \Gamma_d t)}
{\left (1+\zeta^2_d \right ) + \left (1-\zeta^2_d \right )
\cos (x_d \Gamma_d t) - 2 \zeta_d \cos (2\beta)
\sin\delta_d \sin (x_d \Gamma_d t)} \; \; ,
\end{equation}
and
\begin{equation}
{\cal A}_s(t) \; =\; \frac{-2\zeta_s \sin (2\beta')
\cos\delta_s \sin (x_s \Gamma_s t)}
{\left (1+\zeta^2_s \right ) + \left (1-\zeta^2_s \right )
\cos (x_s \Gamma_s t) - 2 \zeta_s ~ \cos (2\beta')
\sin\delta_s \sin (x_s \Gamma_s t)} \; \; .
\end{equation}
Clearly these two formulas would be simplified to the
familiar form, if $\zeta_{d(s)} =1$ and $\delta_{d(s)}
=0$ held (i.e., if the final states were $CP$ eigenstates).
A necessary condition for large $CP$ violation is
that $\zeta_d$ or $\zeta_s$ does not deviate much from
one. Subsequently one can see that this condition is
indeed satisfied -- both $\zeta_d$ and $\zeta_s$ are of 
$O(1)$, estimated in the factorization approximation.

\section{Factorization and penguin effect}

Now we calculate the decay amplitudes of 
$B^0_d \rightarrow D^{*\pm}D^{\mp}$ and 
$B^0_s \rightarrow D^{*\pm}_s D^{\mp}_s$ by use
of the effective weak Hamiltonian and the factorization
approximation. The contributions induced by the 
annihilation-type quark diagrams at the tree and penguin
levels are expected to have significant form-factor
suppression and will be neglected. The relevant $\Delta B =1$
Hamiltonian can then be written as \cite{Buras}
\begin{equation}
{\cal H}_{\rm eff} \; =\; \frac{G_{\rm F}}{\sqrt{2}}
\left [ (V_{cb}V^*_{cq}) \sum^2_{i=1} (c_i ~ O^c_i)
~ - ~ (V_{tb}V^*_{tq}) \sum^{10}_{i=3} (c_i ~ O_i)
\right ] \; ~ + ~ {\rm h.c.} \; ,
\end{equation}
where $q=d$ or $s$, $c_i$ (for $i=1, \cdot\cdot\cdot ,10$)
are the Wilson coefficients, and
\begin{eqnarray}
O^c_1 & = & (\bar{c}b)_{V-A} (\bar{q}c)_{V-A} \; , 
~~~~~~~~~~~~~~~~~~~
O^c_2 \; =\; (\bar{q}b)_{V-A} (\bar{c}c)_{V-A} \; , 
\nonumber \\
O_3 & = & (\bar{q}b)_{V-A} \sum_{q'} (\bar{q'}q')_{V-A} \; ,
~~~~~~~~~~~~~ O_4 \; =\; (\bar{q}_\alpha b_\beta)_{V-A}
\sum_{q'} (\bar{q'}_\beta q'_\alpha)_{V-A} \; ,
\nonumber \\
O_5 & = & (\bar{q}b)_{V-A} \sum_{q'} (\bar{q'}q')_{V+A} \; ,
~~~~~~~~~~~~~ O_6 \; =\; (\bar{q}_\alpha b_\beta)_{V-A}
\sum_{q'} (\bar{q'}_\beta q'_\alpha)_{V+A} \; ,
\nonumber \\
O_7 & = & \frac{3}{2} (\bar{q}b)_{V-A} \sum_{q'}
\left [e^{~}_{q'} (\bar{q'} q')_{V+A} \right ] \; ,
~~~~ O_8 \; =\; \frac{3}{2} (\bar{q}_\alpha b_\beta)_{V-A}
\sum_{q'} \left [e^{~}_{q'} (\bar{q'}_\beta q'_\alpha)_{V+A}
\right ] \; , \nonumber \\
O_9 & = & \frac{3}{2} (\bar{q}b)_{V-A} \sum_{q'}
\left [e^{~}_{q'} (\bar{q'} q')_{V-A} \right ] \; ,
~~~ O_{10} \; =\; \frac{3}{2} (\bar{q}_\alpha b_\beta)_{V-A}
\sum_{q'} \left [e^{~}_{q'} (\bar{q'}_\beta q'_\alpha)_{V-A}
\right ] \; .
\end{eqnarray}
Here $O_3, \cdot\cdot\cdot ,O_6$ denote the QCD-induced penguin
operators, $O_7, \cdot\cdot\cdot ,O_{10}$ denote the electroweak
penguin operators, and $\alpha$ and $\beta$ are the $SU(3)$
color indices. To calculate the physical amplitude of an exclusive
$B$ decay mode, the Wilson coefficients and the
relevant hadronic matrix element of 
four-quark operators need be evaluated in the same 
renormalization scheme and at the same energy scale.
After this procedure \cite{Fleischer}
we are left with the effective Wilson 
coefficients $c^{\rm eff}_i$, and they will enter 
the factorized decay amplitude in the following
combinations \cite{Cheng98}:
\begin{equation}
a^{\rm eff}_{2i} \; =\; c^{\rm eff}_{2i} ~ + ~
\frac{c^{\rm eff}_{2i-1}}{N^{\rm eff}_c} \; \; ,
~~~~~~~~
a^{\rm eff}_{2i-1} \; =\; c^{\rm eff}_{2i-1} ~ + ~
\frac{c^{\rm eff}_{2i}}{N^{\rm eff}_c} \;\; ,
\end{equation}
(for $i=1, \cdot\cdot\cdot , 5$), where $N^{\rm eff}_c$
is the effective number of colors. The values of $a^{\rm eff}_i$
for differrent $N^{\rm eff}_c$ can be found in
Refs. \cite{Cheng98,Lue}.

It is then straightforward to calculate the
decay amplitudes of $B^0_d\rightarrow D^{*\pm}D^{\mp}$,
$B^0_s\rightarrow D^{*\pm}_sD^{\mp}_s$ and their
$CP$-conjugated processes in the 
factorization scheme. For exmaple,
\begin{eqnarray}
A(B^0_d\rightarrow D^{*-}D^+) & = & 
\left \{ (V^*_{cb}V_{cd})
a^{\rm eff}_1 - (V^*_{tb}V_{td} ) \left [
\left (a^{\rm eff}_4 + a^{\rm eff}_{10} \right )
- \left (a^{\rm eff}_6 + a^{\rm eff}_8 \right ) z_d
\right ] \right \} M^{D^+D^{*-}}_{cdc} \; , \nonumber \\
A(B^0_d\rightarrow D^{*+}D^-) & = &
\left \{ (V^*_{cb}V_{cd}) 
a^{\rm eff}_1 - (V^*_{tb}V_{td} ) \left
(a^{\rm eff}_4 + a^{\rm eff}_{10} \right ) \right \} 
M^{D^{+*}D^-}_{cdc} \; , 
\end{eqnarray}
and
\begin{eqnarray}
A(B^0_s\rightarrow D^{*-}_sD^+_s) & = & 
\left \{ (V^*_{cb}V_{cs})
a^{\rm eff}_1 - (V^*_{tb}V_{ts} ) \left [
\left (a^{\rm eff}_4 + a^{\rm eff}_{10} \right )
- \left (a^{\rm eff}_6 + a^{\rm eff}_8 \right ) z_s
\right ] \right \} M^{D^+_sD^{*-}_s}_{csc} \; , \nonumber \\
A(B^0_s\rightarrow D^{*+}_sD^-_s) & = &
\left \{ (V^*_{cb}V_{cs}) 
a^{\rm eff}_1 - (V^*_{tb}V_{ts} ) \left
(a^{\rm eff}_4 + a^{\rm eff}_{10} \right ) \right \} 
M^{D^{+*}_sD^-_s}_{csc} \; , 
\end{eqnarray}
in which
\begin{eqnarray}
M^{XX^*}_{q^{~}_1q^{~}_2q^{~}_3} & \equiv & \frac{G_{\rm F}}{\sqrt{2}}
~ \langle X|(\bar{q}^{~}_1 q^{~}_2)_{V-A}|0\rangle \langle X^*|
(\bar{b}q^{~}_3)_{V-A}|B^0_q\rangle \; \nonumber \\
& = & \sqrt{2} G_{\rm F} ~ m^{~}_{X^*} ~ f_X ~ A_0^{BX^*}(m^2_X)
~ \left (\varepsilon^{~}_{X^*} \cdot k_{B_q} \right ) \; ,  
\nonumber \\
M^{X^*X}_{q^{~}_1q^{~}_2q^{~}_3} & \equiv & \frac{G_{\rm F}}{\sqrt{2}}
~ \langle X^*|(\bar{q}^{~}_1 q^{~}_2)_{V-A}|0\rangle \langle X|
(\bar{b}q^{~}_3)_{V-A}|B^0_q\rangle \; \nonumber \\
& = & \sqrt{2} G_{\rm F} ~ m^{~}_{X^*} ~ f_{X^*} ~ 
F_1^{BX}(m^2_{X^*})
~ \left (\varepsilon^{~}_{X^*} \cdot k_{B_q} \right ) \; ,  
\end{eqnarray}
are the factorized hadronic matrix elements expressed
by relevant decay constants and form-factors, and
\begin{equation}
z_q \; \equiv \; \frac{2m^2_X}{(m_1 + m_2) (m_b + m_3)}
\end{equation}
arises from transforming the $(V-A) (V+A)$ currents into
the $(V-A) (V-A)$ ones for the penguin amplitudes.
In Eqs. (20) and (21), $k_{B_q}$ denotes the momentum of 
$B_q$ meson ($q=d$ or $s$), and $m_i$ is the mass of $q_i$ quark
($i=1,2,3$).

With the help of Eqs. (18) -- (21) we are able to estimate
the penguin contributions to the interference quantities 
$\lambda_{D^{*+}D^-}$ and $\lambda_{D^{*+}_sD^-_s}$ given in
Eqs. (4) and (5), respectively. As $|a^{\rm eff}_1|
\gg |a^{\rm eff}_i|$ (for $i=3, \cdot\cdot\cdot , 10$) holds
\cite{Cheng98,Lue}, some analytical approximations can reliably
be made. To the next-to-leading order we obtain
\begin{eqnarray}
|\lambda_{D^{*+}D^-}| & = & \left [ 1 ~ - ~ 
\left | \frac{V_{tb}V_{td}}{V_{cb}V_{cd}} \right |
\frac{(a^{\rm eff}_6 + a^{\rm eff}_8) z_d}{a^{\rm eff}_1} 
\cos\beta \right ]
\frac{M^{D^+D^{*-}}_{cdc}}{M^{D^{*+}D^-}_{cdc}} \; \; ,
\nonumber \\
|\lambda_{D^{*+}_sD^-_s}| & = & \left [ 1 ~ - ~ 
\left | \frac{V_{tb}V_{ts}}{V_{cb}V_{cs}} \right |
\frac{(a^{\rm eff}_6 + a^{\rm eff}_8) z_s}{a^{\rm eff}_1} 
\cos\beta' \right ]
\frac{M^{D^+_sD^{*-}_s}_{csc}}{M^{D^{*+}_sD^-_s}_{csc}} \; \; .
\end{eqnarray}
Numerically one finds $|(V_{tb}V_{tq})/(V_{cb}V_{cq})|
\approx 1$ for both $q=d$ and $q=s$ cases \cite{PDG98}.
In addition, 
$|(a^{\rm eff}_6 + a^{\rm eff}_8)/a^{\rm eff}_1|
\approx |a^{\rm eff}_6/a^{\rm eff}_1| \approx 5\%$
\cite{Cheng98,Lue} and
$z_d \approx z_s \approx 0.7$. Therefore 
the penguin contributions to $|\lambda_{D^{*+}D^-}|$
and $|\lambda_{D^{*+}_sD^-_s}|$ are expected to
be about $4\%$ or smaller 
for arbitrary values of $\beta$ and $\beta'$.
Neglecting the penguin effects in such decay modes
turns out to be a good
approximation, leading to the following instructive
result:
\begin{eqnarray}
|\lambda_{D^{*+}D^-}| \; =\; 
\zeta_d & = & \frac{M^{D^+D^{*-}}_{cdc}}{M^{D^{*+}D^-}_{cdc}}
\; =\; \frac{f_D}{f_{D^*}} \cdot \frac{A_0^{B_dD^*}(m^2_D)}
{F^{B_dD}_1(m^2_{D^*})} \;\; , \nonumber \\
|\lambda_{D^{*+}_sD^-_s}| \; =\;
\zeta_s & = & \frac{M^{D^+_sD^{*-}_s}_{csc}}{M^{D^{*+}_sD^-_s}_{csc}}
\; =\; \frac{f_{D_s}}{f_{D^*_s}} \cdot \frac{A_0^{B_sD^*_s}(m^2_{D_s})}
{F^{B_sD_s}_1(m^2_{D^*_s})} \;\; .
\end{eqnarray}
As we have emphasized in Eq. (3), $\zeta_d$ and $\zeta_s$
are pure functions of decay constants and form-factors.
Under $SU(3)$ symmetry $\zeta_s = \zeta_d$ holds.
The measurement of these two parameters will 
provide a clean test of the factorization hypothesis.

At present it remains difficult to make reliable
evaluation of the form-factors appearing in Eq. (3) or (23).
From current data one can determine the ratio
$f_{D^*_s}/f_{D_s} = 1.16 \pm 0.19$ \cite{Stech}, but
little experimental knowledge about $f_{D^*}/f_D$ is
now available. For self-consistency
we just use the model-dependent results of
decay constants and formfactors in Ref. \cite{Gatto}
to illustrate the ballpark values of 
$\zeta_d$ and $\zeta_s$. We obtain
$\zeta_d \approx 1.04$ and $\zeta_s \approx 1.03$. 
While these two numbers might not be trustworthy, they
do reflect the qualitative expectation $\zeta_d \sim
\zeta_s \sim 1$. In some sense the final states
$D^{*\pm}D^{\mp}$ and $D^{*\pm}_sD^{\mp}_s$ may be
treated as ``quasi''-$CP$ eigenstates, if the strong
phases $\delta_d$ and $\delta_s$ are not significantly
large. This feature makes these decay modes 
as interesting as $B_d \rightarrow D^+D^-$, $D^{*+}D^{*-}$ and
$B_s \rightarrow D^+_sD^-_s$, $D^{*+}_sD^{*-}_s$ for the
study of $CP$ violation. 

The result $\zeta_d \sim 1$ implies that the $CP$ asymmetry
${\cal A}_d(t)$ in Eq. (13) can approximately be as large
as $\sin (2\beta) \cos\delta_d$ in magnitude. Therefore 
we are able to get a significant $CP$-violating signal
in $B_d\rightarrow D^{*\pm}D^{\mp}$ (e.g., for
$|\cos\delta_d| >0.5$) within the standard model.
In comparison, the $CP$ asymmetry in
$B_s\rightarrow D^{*\pm}_sD^{\mp}_s$ transitions
are vanishingly small in the standard model.
 
\section{Further discussions}

We have discussed $CP$ asymmetries and penguin effects
in $B_d\rightarrow D^{*\pm}D^{\mp}$ and $B_s\rightarrow
D^{*\pm}_sD^{\mp}_s$ transitions. A possibility to 
test the factorization hypothesis for such neutral-$B$
decays has also been pointed out. Before ending this
work we make two useful comments.

(1) Beyond the standard model, new physics may introduce
an additional $CP$-violating phase into $B^0_d$-$\bar{B}^0_d$
or $B^0_s$-$\bar{B}^0_s$ mixing. In this case the weak
phases to be measured from $B_d\rightarrow D^{*\pm}D^{\mp}$
and $B_s\rightarrow D^{*\pm}_sD^{\mp}_s$ decay modes will
not be $\beta$ and $\beta'$. For illustration we only consider
the kinds of new physics that do not violate unitarity of the
CKM matrix and have insignificant effects on the penguin channels
of the above-mentioned decays \cite{Nir}. Then 
$B^0_d$-$\bar{B}^0_d$ and $B^0_s$-$\bar{B}^0_s$ mixing phases
can be written as
\begin{equation}
\frac{q^{~}_d}{p^{~}_d} \; =\; \frac{V^*_{tb}V_{td}}{V_{tb}
V^*_{td}} e^{{\rm i}\phi_{\rm NP}} \; , ~~~~~~~~
\frac{q^{~}_s}{p^{~}_s} \; =\; \frac{V^*_{tb}V_{ts}}
{V_{tb}V^*_{ts}} e^{{\rm i}\phi'_{\rm NP}} \; ,
\end{equation}
in which $\phi_{\rm NP}$ and $\phi'_{\rm NP}$ denote
the $CP$-violating phases induced by new physics. The 
overall weak phases of $\lambda_{D^{*+}D^-}$ and
$\lambda_{D^{*+}_sD^-_s}$ turn out to be
\begin{equation}
2\beta \; \longrightarrow \; 2\beta + \phi_{\rm NP} \; ,
~~~~~~~~
2\beta' \; \longrightarrow \; 2\beta' + \phi'_{\rm NP} \; ,
\end{equation}
respectively. Therefore only the combinations 
$(2\beta + \phi_{\rm NP})$ and $(2\beta' + \phi'_{\rm NP})$,
instead of $\beta$ and $\beta'$ themselves, can be extracted
from the proposed time-dependent measurements.

In principle it is possible to isolate
$\phi_{\rm NP}$ and $\phi'_{\rm NP}$ from the decay modes
$B_d\rightarrow K^0\bar{K}^0$ and $B_s\rightarrow K^0\bar{K}^0$
respectively, which occur only through the penguin-induced
quark diagrams. Assuming the top-quark dominance, one 
obtains the relevant interference quantities between decay
amplitudes and $B^0$-$\bar{B}^0$ mixing as follows:
\begin{eqnarray}
{\rm Im} \left [ \frac{q^{~}_d}{p^{~}_d} \cdot
\frac{A(\bar{B}^0_d\rightarrow K^0\bar{K}^0)}
{A(B^0_d\rightarrow K^0\bar{K}^0)} \right ] & = &
{\rm Im} \left ( \frac{q^{~}_d}{p^{~}_d} \cdot
\frac{V_{tb}V^*_{td}}{V^*_{tb}V_{td}} \right )
\; =\; \sin\phi_{\rm NP} \; , \nonumber \\
{\rm Im} \left [ \frac{q^{~}_s}{p^{~}_s} \cdot
\frac{A(\bar{B}^0_s\rightarrow K^0\bar{K}^0)}
{A(B^0_s\rightarrow K^0\bar{K}^0)} \right ] & = &
{\rm Im} \left ( \frac{q^{~}_s}{p^{~}_s} \cdot
\frac{V_{tb}V^*_{ts}}{V^*_{tb}V_{ts}} \right )
\; =\; \sin\phi'_{\rm NP} \; .
\end{eqnarray}
Unfortunately these decays have quite small
branching ratios: ${\cal B}(B^0_d\rightarrow
K^0\bar{K}^0) \sim 10^{-6}$ \cite{Lue} and
${\cal B}(B_s^0\rightarrow K^0\bar{K}^0)
\sim 10^{-5}$ \cite{Cheng98}. Hence they could
only be meeasured in the second-round experiments
of $B$-meson factories.

(2) It is also worth pointing out that the joint decay
of a coherent $B^0_d\bar{B}^0_d$ pair into 
$D^{*\pm}D^{\mp}$ states on the $\Upsilon (4S)$
resonance, i.e.,
$$
\left (B^0_{d,\rm phys} \bar{B}^0_{d, \rm phys}
\right )_{\Upsilon (4S)} \rightarrow (D^{*+}D^-)_{B_d}
(D^{*-}D^+)_{B_d} \; ,
$$
is interesting, although the decay rate
is considerably suppressed (of order $10^{-8}$). 
Following the general discussions made in Ref. \cite{Bigi91}, we
explicitly obtain this joint decay rate (divided by
$|A(B^0_d\rightarrow D^{*+}D^-)|^4$) as 
\begin{eqnarray}
{\cal R}_{\Upsilon (4S)} 
& \propto & \frac{2+x^2_d}{1+x^2_d} 
\left |\lambda_{D^{*+}D^-} - \frac{1}{\bar{\lambda}_{D^{*-}D^+}}
\right |^2 ~ + ~ \frac{x^2_d}{1+x^2_d} 
\left | 1- \frac{\lambda_{D^{*+}D^-}}
{\bar{\lambda}_{D^{*-}D^+}} \right |^2 \nonumber \\
& = & \frac{2 + x^2_d}{1 + x^2_d} \left [
\zeta^2_d + \frac{1}{\zeta^2_d} 
- 2 \cos (\delta_d) \right ] 
~ + ~ \frac{4 x^2_d}{1 + x^2_d} \sin^2 (2\beta)  \; .
\end{eqnarray}
We observe that the $CP$-violating contribution to the
decay rate is significant and could finally be separated out.
Such a measurement might be available in the second-round
experiments of $e^+e^-$ $B$-meson factories with about 
$10^{9-10}$ $B^0_d\bar{B}^0_d$ events.

Our conclusion is simply that the decay modes 
$B_d\rightarrow D^{*\pm}D^{\mp}$ and $B_s\rightarrow 
D^{*\pm}_sD^{\mp}_s$ are very useful for the
study of $CP$ violation and for the test of factorization.
Some special attention is worth being
paid to them in the upcoming
experiments of $B$ physics.

\vspace{0.5cm}
\begin{flushleft}
{\Large\bf Acknowledgement}
\end{flushleft}

The author likes to thank H.J. Lipkin for useful
discussions on this topic during NOW'98 in Amsterdam.

\end{document}